# The mechanism of nanoparticle precipitation induced by electron irradiation in transmission electron microscopy

N. Jiang

Department of Physics, Arizona State University, Tempe, AZ 85287-1504, USA

Ru nanoparticles (NPs) can precipitate in Ru doped $SiO_2$ amorphous thin films, triggered by electron irradiation in (scanning) transmission electron microscope ((S)TEM). A new mechanism was introduced to interpret the formation of metal NPs in (S)TEM. The induced electric field by electron irradiation, which originates from charging due to ionizations and excitations of atom electrons, can reduce the Gibbs free energy barrier for nucleation of metal particles. Furthermore, the directional ion drifting driven by the electric forces may accelerate the kinetic process of metal particle precipitation.

1. Introduction

Nanoparticle (NP) precipitation triggered by electron irradiation is a phenomenon often observed in (scanning) transmission electron microscopy ((S)TEM) studies [1 – 6]. Although this is a type of unwanted specimen damages in pristine materials [4, 5], sometimes this may serve as a synthetic tool to produce nanostructure in desired materials [6]. Either to avoid or to use it, one needs to understand mechanisms to optimize experimental conditions as required. Previously, both knock-on and radiolysis have been overwhelmingly considered as the main causes of NP precipitation or beam damage [4, 6]. The former is due to kinetic energy and momentum transfer from a beam electron to an atom, resulting in displacement of the affected atom. The latter is originated from ionizing atom electrons by beam electrons, and decays of the excited electrons may result in displacement of an atom (specifically an anion) [for a review see 5]. Apparently, these two mechanisms can easily scramble an ordered structure into a disordered state by randomly displacing atoms, but it is not obvious why they can also drive the randomly displaced atoms into a new and ordered structure (or phase). Temperature rise due to the energy deposit may be a possibility [7], but most experimental evidences strongly suggest that the precipitation be a result of an athermal crystallization process [3, 8–10]. Thermodynamically, the precipitation (or phase transformation) is driven by lowering the Gibbs free energy of system. However, in (S)TEM, electron irradiation continuously injects energy into the specimen. To overcome this dilemma, a two-energy-levels model was suggested [11]. A part of energy input may be dissipated into the environment as the irradiated specimen's atoms rearrange to relax the atomic structure. During the rearrangement process the specimen is driven to a stimulated higher-energy state ($E_2$), which is thermodynamically unstable and quickly decays releasing some energy. After the rearrangement is complete the internal energy of the specimen ($E_3$) drops below the original internal energy ($E_1$), i.e. $E_2 > E_1 > E_3$ [11].

In recent studies, the induced electric field by electron irradiation has been identified as the main cause for beam damage in a variety of materials, in which the formation of NPs triggered by electron irradiation is categorized as phase separation [5, 12, 13]. There are several types of NP formations. From the mass-conservative point of view, one type involves drastic mass loss from the beam-irradiated region. In these materials, some species are more volatile than others under electron beam; the induced electric field may liberate them into vacuum or to adjacent region [6]. The species left in the irradiated region and/or ejected to the adjacent region may form particles [14-17]. Usually, these particles are formed on surfaces of specimen or supporting thin films.

Besides, there is another type of precipitation, in which the mass in the irradiated region does not have noticeable change or the mass loss is not directly associated with the NP precipitation [1 – 3]. In this study we only focus on this type of precipitation and demonstrate experimentally the nucleation process in Ru doped $SiO_2$ amorphous films. Several characteristics can be easily recognized, which include random distribution of NPs, and their small sizes. Most importantly, the precipitation can be identified as a nucleation dominated process, of which the usual coalescence and aging stages do not occur. These characters are common in the electron-beam triggered precipitation and have also been often observed in other materials [1]. Furthermore, we provide a detailed explanation why and how electron beam irradiation can produce the NPs in these materials based on the convention nucleation and growth theory under electric field. All experimental observations can be well interpreted by the proposed mechanism of the induced electric field.

2. Experimental

Amorphous Ru-Si-O films were deposited by reactive rf magnetron co-sputtering from a $Ru_{20}Si_{90}$ 25 cm diameter, fine-grained composite target. The sputtering gas was a mixture of 11% oxygen in argon at a total pressure of 10 mTorr. According to previous work this mixture guaranteed that an oxygen-saturated film would be produced [18, 19]. Substrates were 20 nm thick silicon monoxide (SiO) films suspended on 300 mesh copper grids (Ted Pella, Inc), which had been stripped of their formvar protective film by solvent washing. These substrates were kept near room temperature during deposition by mounting them on a massive copper holder. A deposition rate of 0.1 nm/s was achieved at an rf power of 30 watts. The films produced for this study had a nominal thickness of 20 nm.

The Ru NPs were precipitated and analyzed using JOEL 2010F (S)TEM, equipped with Gatan electron energy loss spectrometer, operating at both TEM and STEM illumination mode at 200kV. The precipitation of Ru NPs was observed by *in situ* imaging, electron diffraction and electron energy loss spectroscopy (EELS) techniques. The energy resolution of EELS was about 1.0 eV.

3. Results

The as-deposited Ru-Si-O thin films used in our study are uniform and no pre-existing particles can be observed. The thin films are also stable; there is no morphology change in the samples after being stored in air at atmospheric pressure for more than a year. However,

Ru NPs can be easily precipitated after the thin films are exposed to electron beam in (S)TEM.

Figure 1 shows a time series of phase contrast images of the Ru-Si-O thin film. The first image (initial) was taken right after the area was exposed to electron beam. Overall, the initial image is quite smooth and homogeneous. The fine grainy contrast is due to the high frequency noise of phase contrast. After 4 minutes of exposure to the same beam of electrons, very faint particle-like contrasts start to occur, and they were everywhere in the illuminated region. The contrasts become stronger and stronger thereafter. After 12 minutes of exposure, the crystal structure can be seen in these particles. The particle sizes do not increase significantly with further electron beam exposure. The average size is about 3.5 nm after 12 minutes of exposure, while it only increases slightly to about 4.0 nm after 23 minutes of exposure. Overall, the sizes of these nanoparticles are in the range of 3 – 5 nm. Figure 2 is a higher magnification image recorded after the particles have been well precipitated by the electron beam. The crystalline lattices of nanoparticles can be easily seen.

The electron diffraction patterns corresponding to different stages of exposure are given in Fig. 3. The initial diffraction shows that the thin film was amorphous (Fig. 3a): no sharp diffraction spots or rings can be seen. With the increase of exposure to electron beam, diffraction rings occur, and their intensities gradually increase with time (Figs. 3b and 3c). After 24 min of exposure, the discontinuity in these rings can be seen clearly (Fig. 3d). The indexes of the visible rings are also given in Fig. 3, which match the calculated diffraction patterns of *hcp* Ru (see S.I.).

The precipitation process can be more easily quantified by time dependent EELS technique. Figure 4 is a time series of Ru $M_{23}$ and O K-edge EELS of the Ru-Si-O thin film. All the spectra were recorded from the same area under the same conditions. The Ru $M_{23}$-edge consists of two peaks at about 463 and 486 eV, corresponding to $M_3$ and $M_2$ peaks respectively. The O K-edge has a major peak at about 539 eV along with a small pre-edge bump at about 533 eV. According to EELS analysis (see S.I.), this small bump in the O K-edge can be assigned to Ru – O bonds. The intensities of Ru $M_{23}$ and O K-edge in this time series were integrated and plotted in Fig. 5. It shows that both Ru and O were losing during the irradiation, but rates are very low, and it is reasonable to assume that this is a mass-conservative system.

The time-dependent changes of near-edge fine structures in the Ru $M_{23}$-edge are not dramatic, but a closer comparison indicates that the widths of $M_{23}$ peak become narrower with the increase of irradiation, and its position shifts about 1.0 eV lower after a long exposure (See S.I.). These changes indicate the transition of Ru from its ionic to metallic state. Such a transition is more profound in the change of the Ru – O peak in the O K-edge, as shown in Fig. 5. It is noticed that the Ru – O peak does not change in a certain period of initial irradiation, about 30 ~ 40 seconds in this case. This can be identified as an incubation period, during which nucleation does not occur. Continuing irradiation causes the peak to decrease, indicating the transition of Ru valence state. As shown in Fig. 5, the peak drops almost linearly to the one-third of its initial intensity within about 200 seconds of irradiation. We consider this period as a nucleation stage, during which some of Ru ions are neutralized and assembled into NPs. Further longer irradiation, however, the transition rate becomes very low, and the most remaining Ru ions remain in ionic state. Since discrete diffraction patterns are evolved from the continuous rings in this period, we can consider it as a ripen stage, in which the degree of crystallinity of Ru NPs increases and lattice images can be easily observed.

Apparently, these stages can be also identified from the overall changes of Ru and O during the irradiation (Fig. 5 and Table I). In the incubation period, both Ru and O remain the same, along with the Ru – O bonds. Their decreasing rates, although very low (Table I), are relatively high in the nucleation stage compared to those in the ripen stage. In fact, the rate at which Ru decreases becomes almost zero in the latest stage. All these phenomena will be interpreted in the following section.

4. Discussion

Thermal annealing of oxygen-saturated Ru-Si-O thin films at 800°C in vacuum can precipitate $RuO_2$ crystals [18, 19]. The electron diffractions, however, can rule out the formation of Ru oxide in this study. This indicates that a dramatic rise in temperature in the Ru-Si-O thin films by electron beam is unlikely to happen. Therefore, the thermal effect alone can be ignored in the precipitation of Ru NPs by electron beam.

The thermodynamic driving force for first-order phase transition is the gain in free energy upon nucleation, the quantity known as supersaturation. In homogeneous nucleation, the total Gibbs free energy, $\Delta G$, to precipitate a NP in a matrix material is the sum of the surface free energy and the bulk free energy. The energy barrier for nucleation, $\Delta G^*$, is the maximum of $\Delta G$, with respect to the size of a nucleus. The driving force is needed to overcome this energy barrier for the formation of new phase. Although the thermal effect may not be the main cause for the Ru NPs precipitation, the mechanism should be able to fulfil this energy requirement.

Apparently, neither knock-on nor radiolysis could suggest the system to overcome the nucleation barrier to precipitate Ru NPs rapidly and massively under electron irradiation. The knock-on interaction results in energy and momentum transfer from an energetic electron to an atom following collision. To displace an atom, the minimum kinetic energy of beam electrons is required. The threshold beam energy is proportionally dependent on the atomic mass and its threshold displacement energy. Considering that the threshold surface sputtering energy is only a fraction of bulk displacement energy and thin specimen for (S)TEM experiments, knock-on interaction is more likely to cause preferential surface sputtering of lighter O and Si than the precipitation of heavier Ru. However, the EELS analysis shows that there is no significant change in both Si and O during the precipitation. Although radiolysis has been widely referred to in terms of beam effects in (S)TEM for decades, the processes are unfortunately poorly described. Apparently, all the detailed processes involve the displacements of negative ions only, which cannot directly explain the rapid assembly of a vast amount of Ru cations into nanoparticles within a short period of time.

As we knew, electric field carrier energy and can affect the free energy barrier of nucleation [20]. According to recent studies [5], strong electric fields can be induced in insulating specimen by electron irradiation in (S)TEM. The electric field is produced by the accumulated charges. These charges are not directly from the beam electrons in transmission geometry. The charging process is initiated by the excitations of atomic electrons. The holes left by the emissions cannot be neutralized in a certain period, resulting in the accumulation of positive charges and in turn they generate electric fields [12, 13, 21, 22].

In the presence of external electric field, the Gibbs free energy is

$$\Delta G = -V \cdot \Delta g + S \cdot \sigma + G_E \qquad (1)$$

Here the first two terms are Gibbs free energy in the absence of electric field and $G_E$ is the change in electrostatic energy due to the formation of a NP. In eqn. (1), $\Delta g$ is the term that includes both the difference in chemical potentials of matrix and NP at the corresponding minima and strain energy generated when a NP forms, and $\sigma$ is the surface tension, and V and S are, respectively, volume and surface area of NP. The influence of an external electric field on the nucleation process has been well studied experimentally and theoretically [23 – 27]. Under the assumption that the precipitated NP is spherical, and the electric field **E** has uniform strength, eqn. (1) has the form of [20]

$$\Delta G = -V \cdot (\Delta g + c_\varepsilon E^2) + S \cdot \sigma \qquad (2)$$

Here the parameter $c_\varepsilon$ is defined as

$$c_\varepsilon = 3\varepsilon_0 \varepsilon_m (\varepsilon_c - \varepsilon_m)/2(\varepsilon_c + 2\varepsilon_m) \qquad (3)$$

in which $\varepsilon_0$ is permittivity of free space, and $\varepsilon_m$ and $\varepsilon_c$ are, respectively, the dielectric constant of matrix and NP. Since $c_\varepsilon$ is NP size independent, $\Delta g + c_\varepsilon E^2$ can be considered as an effective supersaturation, which is a function of the strength of electric field. If $\varepsilon_c > \varepsilon_m$ (NP has higher dielectric constant than the matrix), the electric field can increase the effective supersaturation. If $\varepsilon_c < \varepsilon_m$, it opposites. For metal NPs, their dielectric constants can be considered as infinite, so $c_\varepsilon \approx 1.5\varepsilon_0\varepsilon_m$. Therefore, from free energy point of view, the electric field is always in favor of metal NP precipitation.

For a spherical NP in a uniform electric field, the energy barrier and the critical radius of nucleus are respectively $\Delta G^* = \frac{16\pi}{3} \frac{\sigma^3}{(\Delta g + c_\varepsilon E^2)^2}$ and $\Delta r^* = \frac{2\sigma}{\Delta g + c_\varepsilon E^2}$. Therefore, electron-beam effect on nucleation can be evaluated if $\Delta g$ and $\sigma$ are known. As an approximation, we can estimate $\Delta g$ using $\Delta g(T) = H_f \frac{T_m - T}{T_m} \frac{7T}{T_m + 6T}$, in which $H_f$ and $T_m$ are heat of fusion and melting temperature of metal NP, respectively [28]. The surface energy is set as $\sigma = 0.3$ J/m$^2$ [29]. For Ru, $H_f = 24$ kJ/mole and $T_m = 2337$°C. Suppose $\varepsilon_m = 4.0$ for Ru doped SiO$_2$, the calculated $\Delta G^*$ and $\Delta r^*$ versus electric field strength are plotted in Fig. 6. It is seen that for weak induced electric fields, electron-beam almost has not effect on nucleation of a new phase. However, if the induced electric fields are strong, e.g. close to $10^9$ V/m, both the energy barrier and the critical radius of nucleus are largely reduced. If the fields are even stronger, i.e. > $10^{10}$ V/m, it seems that the nucleation of metal phases becomes an effortless process, from the energy point of view. In our previous studies, it was found that the induced electric fields in SiO$_2$ are in the magnitudes of $10^9 \sim 10^{11}$ V/m [5, 22]. Therefore, this mechanism can explain the rapid nucleation of Ru metal NPs induced electron irradiation without a thermal assistant.

Although the electric field can lower the Gibbs free energies (if $\varepsilon_c > \varepsilon_m$), the precipitation under electron beam is not as the same as that by thermal annealing. Neither RuO$_2$ nor ruthenium silicides is the product of electron irradiation, considering that they all have high conductivities (i.e. $\varepsilon_c > \varepsilon_m$). This is because the strong electric field can also affect the kinetics of the nucleation. One of such factors is atomic drifting under electric fields. Unlike the thermal effect, in which atomic drifting is random, electric forces exerting on ions are directional; forces on positively and negatively charged ions are opposite. According to Ren and Wang [29], the moderate electric fields enhance NaCl nucleation in saturated solutions, but the strong electric fields retard or even prevent the nucleation process of NaCl. This was interpreted as due to the competition between ion self-diffusion and drift motion. The same explanation can be also applied to Ru ions in the silicon oxide thin films. High mobilities of heavy atoms in the Si-O network under electric fields have been observed previously [30]. The strong fields induced by electron irradiation make the oppositely charged ion more difficult to associate, and therefore prevent the formation of $RuO_2$.

Due to the same reason, the electric field can also affect the nucleation rate, by affecting the diffusion coefficient in volume-diffusion control or by changing the mechanism of atom attachment [20]. The threshold character of the dependence of the nucleation rate on the effective supersaturation results in an induction period, which may elapse prior to the formation of a detectable number of NPs. In (S)TEM the induced electric field is not stationary, but increases as the positive charges are building up, until a dynamic equilibrium is reached [5]. Therefore, there should be also a time threshold, before which the field is not strong enough to accelerate or decelerate the nucleation process. In our experiments, the incubation phenomenon was clearly observed in Fig. 5, and its length decreases with the increase of beam current density. During this incubation period, there is evidence that the electric field is not strong enough to move Ru ions (repelling outside the beam region), as shown in Fig 5. This suggests that the induction period of nucleation in (S)TEM may be dominated by the exposure time threshold, which is determined by both the activation energy of Ru migration and beam current density [5].

As we knew [refs], the induced electric fields in (S)TEM are not uniform, and they are strongly dependent on illumination mode and specimen shape. As illustrated in Fig. 7, the strength of electric field is the strongest near the surfaces of specimen and drops rapidly inside the specimen in the center area of the beam in a broad beam TEM illumination. Meanwhile, the fields are also strong near the edge region of the beam. Therefore, we can expect that the spatial distribution of precipitated NPs may not be uniform in the specimen but may preferentially occurs at the sub-surface regions and around edge of electron beam, as illustrated in Fig. 7. In a focused beam STEM illumination, the field has the maximum strength along the probe and drops rapidly away from the probe [12]. Therefore, different from TEM illumination, the precipitation should be three-dimensional and uniform if the STEM probe is scanning across an area.

Dielectric constants of gas phases are usually smaller than that of the matrix materials. From the free energy point of view, the induced electric field is therefore not in favor of nucleation of gas bubbles, which $O_2$ bubbles or clusters were not observed during the Ru precipitation in our experiments. However, the formations of gas bubbles or clusters in some materials are also common phenomena of beam effect [5, and references therein]. In the previous studies, we found that the formation of gas bubbles or clusters is always accompanied by the rapid depletion of cations in the illuminated region [31]. The field-driven cation depletion results in the increase of anion concentration, and thus the supersaturation of anion. Therefore, we speculate that the precipitation of metal NPs is in competition with the formation of gas bubbles in the same region. Bubbles should be more easily observed in materials containing cations with higher mobility, such as alkali and hydrogen, while the metal NPs should be more easily precipitated in materials containing cations with lower mobility, such as transition metal ions. For example, we observe $O_2$ bubbles inside the illuminated region in Na silicate glasses but not Na NPs, which are formed outside the illuminated region [5].

**Conclusions**

A new mechanism of metal NPs precipitation induced by electron irradiation in (S)TEM has been introduced. The induced electric field by electron irradiation, due to ionizations and excitations of atom electrons, can reduce the Gibbs free energy barrier for nucleation of metal particles. Furthermore, the directional ion drifting driven by the electric forces may accelerate the kinetic process of metal particle precipitation.

**Acknowledgement** The experimental work was funded by the NSF DMR-0603993. The amorphous Ru-Si-O films were provided by Dr. G. G. Hembree of ASU. The use of the facility within the center for Solid State Science of ASU is also acknowledged.

Caption

Figure 1   Selected images in a Time series of phase contrast imaging showing the evolution of nanoparticles in the Ru-Si-O thin films by high-energy electron irradiation. The current density of electron beam was 10.4 PA/cm$^2$, and the exposure time for each image was 1 second.

Figure 2   High resolution TEM image showing lattice fringes in individual nanocrystals of Ru.

Figure 3   Time series of electron diffraction patterns showing the crystallization of Ru nanoparticles in the Ru-Si-O thin films.

Figure 4   Selected spectra in a Time series of Ru $M_{23}$ and O K – edge EELS. The acquisition time of each spectrum was 5s under the current density of about 60 PA/cm$^2$. The spectrum indicated as 5 s was the first one, which was acquired immediately after the area was exposed to electron beam.

Figure 5   Integrated intensities of Ru $M_3$ (459 – 469 eV) and O K-edge (529 – 544 eV) in Fig. 4. The energy window for the O pre-edge peak is (529 – 534 eV). All intensities are normalized to their first values in the series.

Figure 6   Diagram demonstrating the effect of electric field on both Gibbs free energy and critical radius of nucleus.

Figure 7   Diagram showing the induced electric field in specimen and the distribution of precipitated NPs.

Table I

Changing rates of the integrated intensities in different stages. They were evaluated using data in Fig. 5 based on a linear regression assumption. Ru – O represents amount of O bound to Ru.

|        | Incubation | Nucleation           | Ripen                |
|--------|------------|----------------------|----------------------|
| Ru     | —          | $-7.4 \times 10^{-4}$ | ~ 0                  |
| O      | —          | $-4.4 \times 10^{-4}$ | $-2.3 \times 10^{-4}$ |
| O – Ru | —          | $-3.2 \times 1.0^{-3}$ | $-5.2 \times 10^{-4}$ |

Figure 1

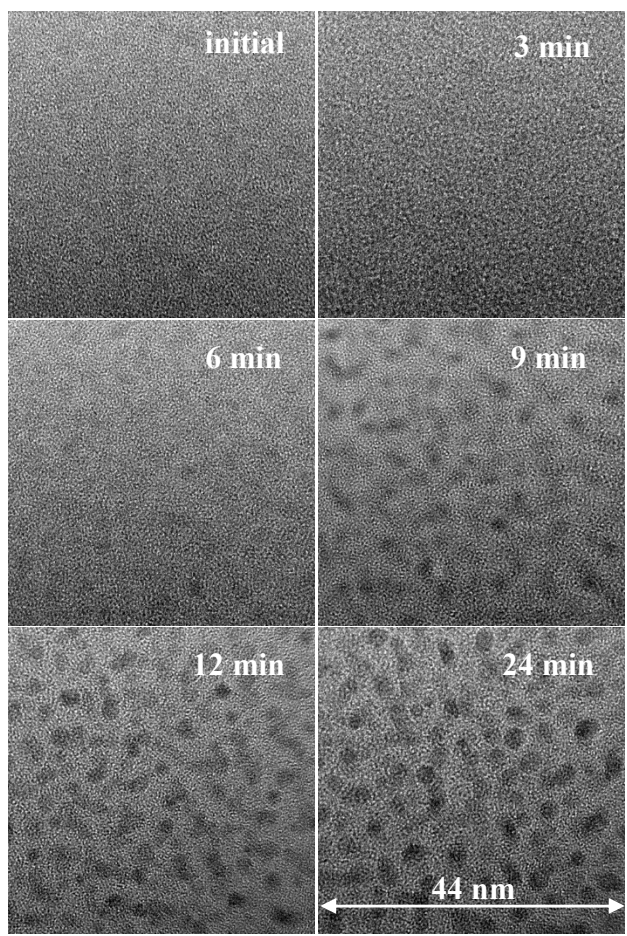

Figure 2

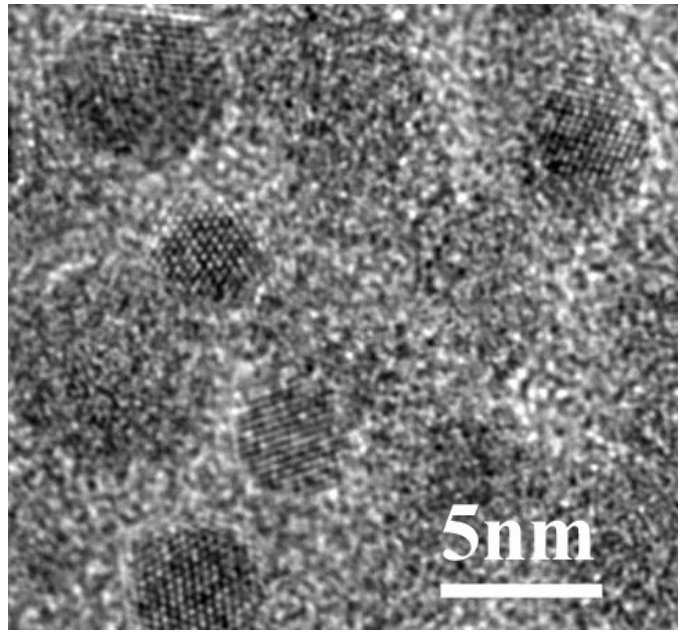

Figure 3

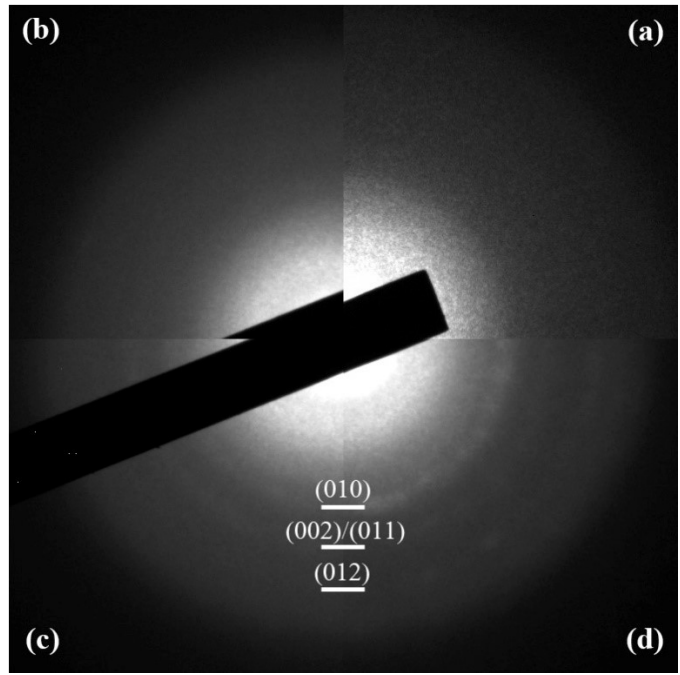

Figure 4

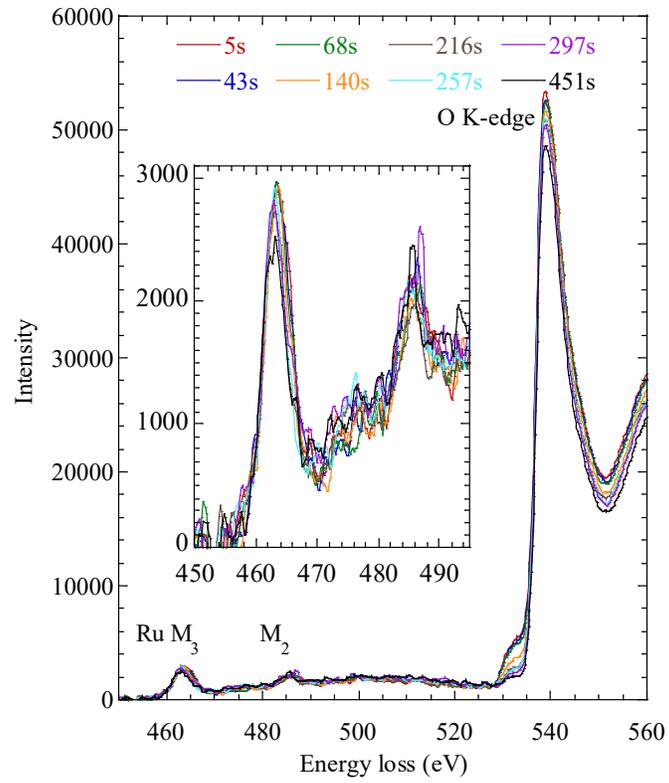

Figure 5

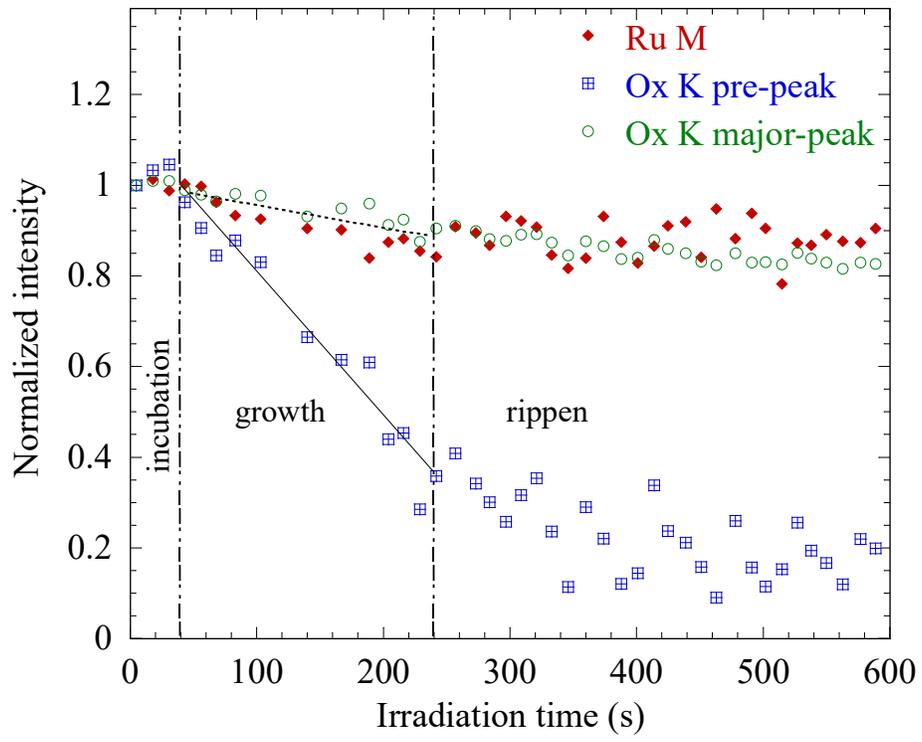

Figure 6

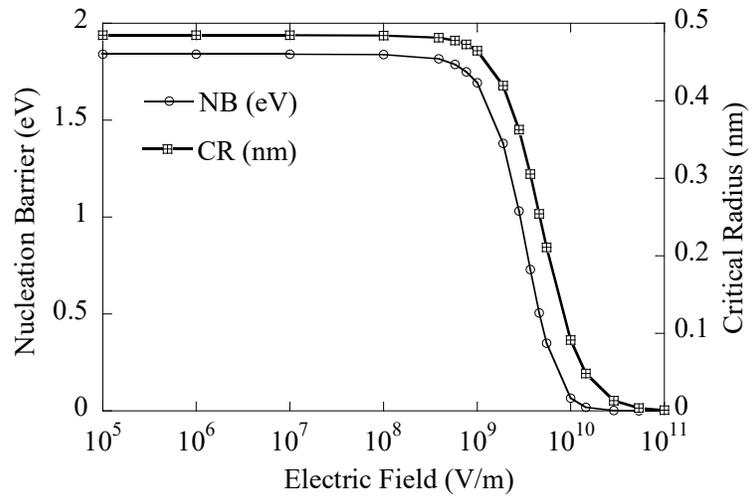

Figure 7

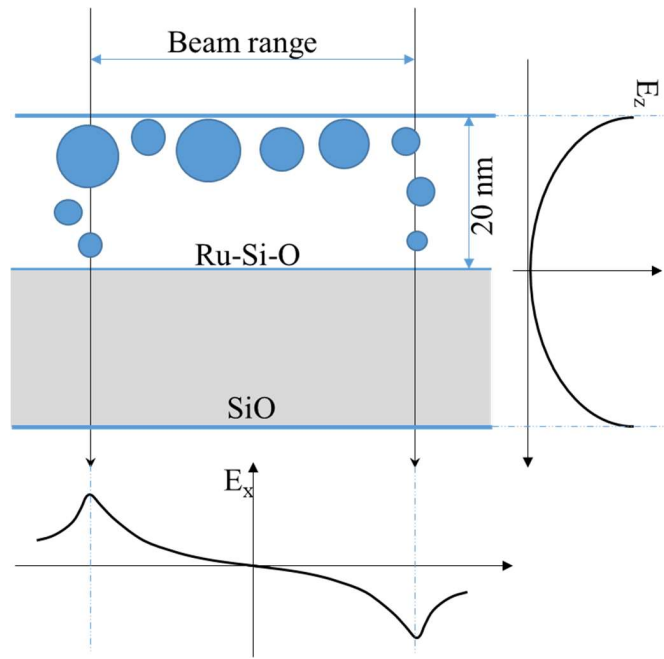

Supplementary Information

Nan Jiang

Department of Physics, Arizona State University, Tempe AZ 85287-1504

The diffraction patterns in Figure 3 were calibrated using a standard Al sample. The measured d-spacing values of the first three visible rings are 2.32, 2.13 and 1.58Å, respectively. These values are very close to the d-spacing values of (010) (2.34Å), (002) (2.14Å), (011) (2.05Å), and (012) (1.58Å) of hcp Ru, respectively. The d-spacing values of (002) and (011) are very close to each other, and thus they cannot be resolved in our measurements because of small particle sizes.

The interpretation of O K-edge of Ru-Si-O thin film can be obtained by the detailed comparisons with the references of SiO and $RuO_2$, which is given in figure S1. In SiO, the O K-edge has only one strong peak at about 539 eV, although there is a very small pre-edge bump. The pre-edge bump of Ru-Si-O is much stronger than that of SiO. On the contrary, the pre-edge bump of Ru-Si-O fits the first two peaks of the O K-edge of $RuO_2$ quite well, as shown in figure S1. So it is reasonable to conclude that that the main features in the Ru-Si-O are dominated by the Si – O characteristic, while the pre-edge peaks can be interpreted as due to the interaction of O – Ru bonds. Two small peaks at 530.5 and 533.0 eV in $RuO_2$ correspond to the "$t_{2g}$" and "$e_g$" peaks, respectively, which are induced by the crystal field of octahedral symmetry around Ru in $RuO_2$ [S1]. The interaction between the Ru d-orbital with the ligand O p-orbitals also results in the same splitting in the O p-DOS, which is responsible for the observed two-peak feature in the O K-edge in $RuO_2$. However, in the Ru-Si-O films, these two peaks cannot be resolved, instead of a broadened bump. This indicates that although Ru atoms or at least some of Ru, in the Ru-Si-O film are bounded to O, the well-organized octahedral coordination as in the $RuO_2$ is unlikely to happen.

The oxidation states of Ru in silicate glasses have not been as well documented as the first row of transition elements, partly because of the low solubility of Ru in the glass perhaps. Nevertheless, a series of optical absorption studies indicated that $Ru^{4+}$ is the most common state of Ru in silicate glasses; while $Ru^{3+}$ and $Ru^{6+}$ also exist [S2]. From structure point of view, Ru is in octahedral coordination in lower valence states (e.g. $Ru^{3+}$ and $Ru^{4+}$), i.e. longer Ru – O distances, while it is in tetrahedral coordination in higher states (e.g. $Ru^{6+}$ and $Ru^{8+}$), i.e. shorter Ru – O distances. In the Ru octahedron, Ru d-DOS splits into two well separated two peaks (~ 2.5 – 3.0eV apart) corresponding to $t_{2g}$ and $e_g$ states, respectively. The calculations show that slight distortions in bond lengths and bond angles in the Ru octahedron do not change the splitting significantly. We therefore believe that the amorphous states may not be the only reason for the broadened pre-edge bump in the O K-edge in the Ru-Si-O system. Instead, the other types of local structure, such as tetrahedral Ru, may also be responsible. The calculations in the tetrahedral Ru, such as RuO4, indicate that the splitting of Ru d- and thus O p-DOS does not occur. This may result in the single broad peak in the O K-edge. This suggestion is also supported by the Ru $M_{23}$ edge in the Ru-Si-O thin film.

As compared in figure S2, the Ru $M_{23}$ edge in the Ru-Si-O does not fully resemble that of $RuO_2$. It is seen that the shape of $M_3$ peak is slightly asymmetry in the Ru-Si-O system, and its position is about 1eV lower than that of $RuO_2$. It is also noted that Ru $M_3$ peak of metal Ru is about 2.5eV lower than that of $RuO_2$ [S3]. On the high-energy side of $M_3$ peak in the Ru-Si-O system, there is a small bump (indicated by an arrow), which is at about the same position as that of $M_3$ peak of $RuO_2$. Piecing together these evidences, it is reasonable to believe that the Ru is in a mixed state between $Ru^{4+}$ (i.e. Ru – O) and some other type of lower valence state(s), such as $Ru^0$.

[S1] Z. Hu, H. von Lips, M. S. Golden, J. Fink, G. Kaindl, F. M. F. de Gtoot, S. Ebbinghaus, and A. Reller, Phys. Rev. B 61, 5262 (2000).

[S2] J. Mukerji and S. R. Biswas, Glass Technology 12, 107 (1971).

[S3] S. P. Kelty, J. Li, J. G. Chen, R. R. Chianelli, J. Ren, and M. –H. Whangbo, J. Phys. Chem. B 103, 4649 (1999).

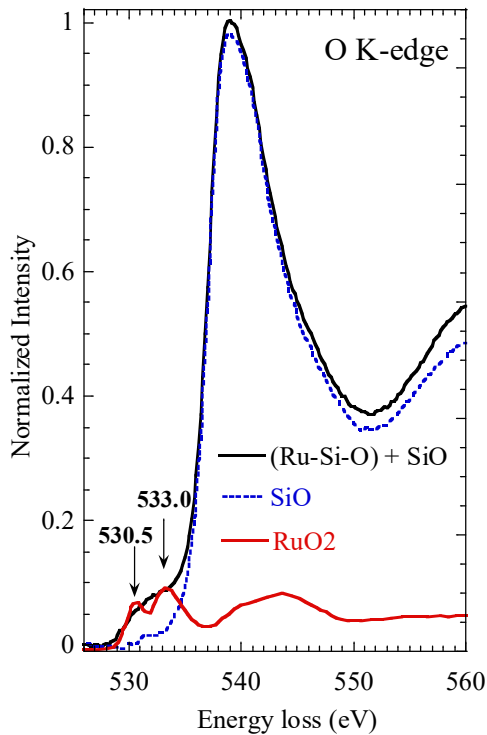

Figure S1. Comparison of O K-edge of Ru – Si – O thin film with Si monoxide and $RuO_2$.

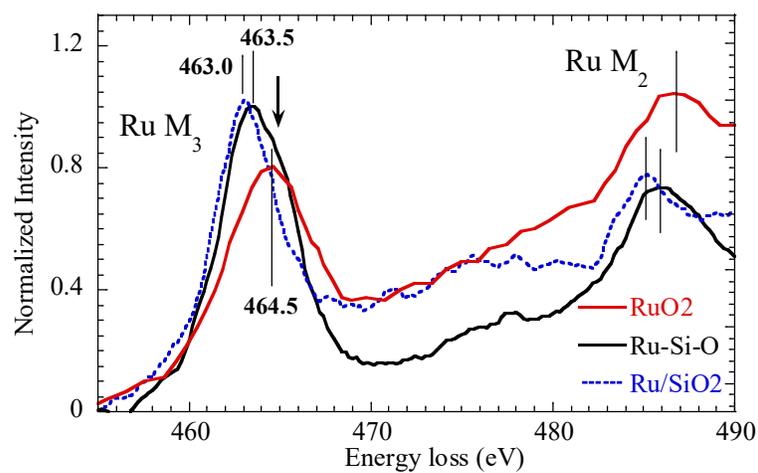

Figure S2. Comparison of Ru $M_{23}$ edge of Ru – Si – O thin film with $RuO_2$. The spectrum marked as $Ru/SiO_2$ was acquired after the precipitation of Ru NPs.